\documentclass[a4paper,11pt]{article}
\pdfoutput=1 % if your are submitting a pdflatex (i.e. if you have
\usepackage{jcappub} % for details on the use of the package, please
\usepackage[T1]{fontenc} % if needed
\usepackage{sidecap}
\usepackage{subcaption}
\usepackage{tabularx,ragged2e,pict2e,booktabs}
\usepackage{array}

\title{Inpainting CMB maps using Partial Convolutional Neural Networks}

\author[a,1]{Gabriele Montefalcone,\note{Corresponding author.}}
\author[b]{Maximilian H. Abitbol,}
\author[b]{Darsh Kodwani}
\author[b]{and R.D.P. Grumitt}

\affiliation[a]{Princeton University, Department of Physics, Jadwin Hall, Washington Road,  Princeton, NJ, 08544, USA}
\affiliation[b]{University of Oxford, Department of Physics, Denys Wilkinson Building, Keble Road, Oxford, OX1 3RH, UK}

\emailAdd{gm9@princeton.edu}
\emailAdd{maximilian.abitbol@physics.ox.ac.uk}
\emailAdd{darsh.kodwani@physics.ox.ac.uk}
\emailAdd{richard.grumitt@physics.ox.ac.uk}

\abstract{We present a novel application of partial convolutional neural networks (PCNN) that can inpaint masked images of the cosmic microwave background. The network can reconstruct both the maps and the power spectra to a few percent for circular and irregularly shaped masks covering up to ~10\% of the image area. By performing a Kolmogorov-Smirnov test we show that the reconstructed maps and power spectra are indistinguishable from the input maps and power spectra at the 99.9\% level. Moreover, we show that PCNNs can inpaint maps with regular and irregular masks to the same accuracy. This should be particularly beneficial to inpaint irregular masks for the CMB that come from astrophysical sources such as Galactic foregrounds. The proof of concept application shown in this paper shows that PCNNs can be an important tool in data analysis pipelines in cosmology.}


\begin{document}
\maketitle
\flushbottom
%%%%%%%%%%%%%%%%%%%%%%%%%%%%%%%%%%%%%%%%%%%%%%%%%%%

\section{Introduction}

Observations of the cosmic microwave background (CMB) have provided crucial experimental tests for theoretical models describing the origin and evolution of the universe. 
We currently have a model that describes the observations from the CMB and other cosmological probes extremely well, called the $\Lambda$CDM model. 
Obtaining precision constraints on cosmological parameters requires a thorough understanding of instrumental and astrophysical systematic errors. 
Of particular importance for CMB measurements, are astrophysical foregrounds; non-CMB sources of microwave emission from our own galaxy and from extra-galactic sources that obfuscate CMB observations~\cite{planck2020compsep}.
Typically, pixels that are completely dominated by foregrounds are removed from the maps and inferences are made from masked maps using cut-sky power spectrum estimators. One of the most widely-used estimators, largely due to their computational efficiency, are pseudo-$C_l$ estimators \cite{Hivon:2001jp, Challinor:2004pr}. Pseudo-$C_l$ methods construct unbiased, but sub-optimal power spectrum estimators on the cut-sky. One may ask whether the missing values can be replaced by physically or statistically motivated values and whether that can allow for better power spectrum estimation than the cut-sky pseudo-$C_l$ estimators. This question has previously been considered in \cite{Gruetjen_2017}, where it was found that using a linear inpainting scheme could significantly outperform pseudo-$C_l$ methods. In this paper we demonstrate the ability of our machine learning based inpainting scheme at recovering the correct input power spectra as a proof of concept. However, this is likely to be particularly advantages when attempting to construct potential non-Gaussian features in masked CMB maps.

The general procedure of replacing missing values through inpainting is a common problem in a variety of fields in addition to cosmology such as computer vision, satellite imaging etc. A review of inpainting algorithms can be found in~\cite{7544859}. 
Machine learning techniques have become ubiquitous in solving inpainting problems as they do not rely on an a-priori model that describes the data. 
Instead they learn statistical features about the data from training data. 
Of course, it is possible that a forward model is required to create a training set in the first place - this is indeed the case for CMB maps, as we only have a single realisation of the CMB map.

Simulating a CMB map requires a set of cosmological parameters that will determine the statistical properties of the map. 
The observed CMB will come from the ``true" cosmological parameters\footnote{This implicitly assumes that the $\Lambda$CDM model is a the correct model of the universe.}. 
Since we do not have access to the true parameters, in order to infer the actual statistical properties of the map we need to sample a range of cosmological parameters and see which one of them best matches the observations. 
The training set we make for inpainting CMB maps includes maps that are evaluated for different cosmological parameters. 

The traditional approach to inpainting missing values is to use a Gaussian constrained realisation of the map \cite{Bucher2012}.
There have been various studies that have recently attempted to use machine learning techniques to inpaint CMB images\footnote{Machine learning techniques have also previously been used more generally in the context of CMB data analysis \cite{ref1,ref2,ref3,ref4,ref5}}. 
In \cite{Yi:2020xgq} a variational auto-encoder is used to reconstruct CMB maps. 
This study compared the difference between the reconstructed image and the original image at the map level, however, in cosmology the more relevant function is the power spectrum as that is the object typically used for inference of cosmological parameters. 
Further studies have used different machine learning methods and also compared power spectra between the true images and the inpainted images. 
In \cite{Puglisi:2020deh} three separate machine learning methods are used to inpaint galactic foreground maps. 
One of the most popular techniques in computer vision is called Generative Adversarial Networks (GANs) \cite{2014arXiv1406.2661G} and it has been used to inpaint CMB maps in \cite{Sadr:2020rje}. 
Moreover, the symmetrical properties of the sphere are taken advantage of by using a neural network architecture that preserves SO(3) symmetry in \cite{Petroff:2020fbf} to inpaint CMB images for foreground removal. 
Traditional deep learning methods use both the masked (empty) pixels and the surrounding pixels with signals in an image to inpaint the masked region. 
This is known to lead to bluriness in the inpainted image \cite{PCNN}.
In this paper we use a novel concept of partial convolutions in convolutional neural networks \cite{PCNN} to inpaint CMB maps. The strength of using partial convolutions is that they only use the non-masked pixels to draw statistical information and inpaint the masked pixels with that information. This is done by masking the convolution matrix itself. Our model outperforms the other recently developed ML methods that inpaint pure CMB maps. Specifically, we obtain
mean square error ($\mathrm{MSE}$), mean absolute error ($\mathrm{MAE}$) and peak signal to noise ratio ($\mathrm{PSNR}$) equal to $0.0005$, $0.0044$ and $32.881$, as opposed to $0.0055$, $0.134$ and $23.989$ obtained in \cite{Yi:2020xgq} for similar mask sizes\footnote{The numerical values for the three parameters presented here correspond to the average values obtained from comparing 500 ground-truth and inpainted CMB patches from the test set and masked with $\mathbf{M_r}$ (see \ref{sec:masking} for more details on this).}. Moreover, we are able to reproduce the CMB patch power spectra to percent accuracy up to $\ell_{\mathrm{max}}=2048$  for irregularly shaped masks that cover $\lesssim 15\%$ of the total area,  compared to $\ell_{\mathrm{max}}\sim1500$ obtained in \cite{Sadr:2020rje} for regular masks of equivalent size.

This paper is structured as follows:
Section \ref{ML} described the machine learning method of partial convolutions that is used for inpainting. 
The results, for the pixel distribution and the power spectra of inpainted maps, are shown in section \ref{results} and finally we conclude in section \ref{conclusion} with a summary of our results and potential future avenues of research using partial convolutions.

\section{Machine learning algorithm
\label{ML}}

Inpainting refers to the process of reconstructing lost or deteriorated parts of an image without introducing significant differences. Several algorithms have been developed to solve such tasks. 
They can be classified into two approaches:
\begin{itemize}
    \item Patch-based synthesis and image statistics \cite{patchmatch,image_stat,imag_stat2,imag_stat3}
    \item Machine learning (ML) techniques involving neural networks \cite{GAN1,CNN1,CNN2,CNN3,CNN4}.
\end{itemize}
We focus on the second approach in this paper.
Convolutional neural networks (CNNs) have proven particularly effective in image analysis, with their network architecture being specifically designed for the extraction of information from image data through a series of convolutional layers.

Each layer consists of a set of learnable filters which are convolved with the input image, sliding across its pixels and identifying key features in the image. 
The deeper the network, the more features that can be extracted, such that, by building a CNN with a sufficiently large number of layers, one is able to recognize even the most intricate images. This, however, can also lead to overfitting and thus has to be accounted for by techniques such as dropout \cite{dropout}. 
Given that the CNN filter responses
are conditioned on both valid pixels as well as the substitute values in the masked holes, a CNN algorithm that fills in
masked parts of an image suffers from the dependence on
the initial hole values. 
This dependency on the hole values can result in several imperfections in the output image such as color discrepancy, lack of texture in the hole regions and blurriness \cite{PCNN}. These artifacts are typically removed by downstream processing of the resulting images which increases the computational cost of inpainting \cite{7968387, 10.1145/882262.882269}. 

In an attempt to solve such inaccuracies in the image reconstruction, whilst minimising the inpainting time and cost, a deep learning algorithm based on partial convolutional\footnote{Sometimes also called \emph{segmentation-aware convolutions}.} neural networks (PCNNs) has been developed in \cite{PCNN, harley2017segmentationaware}. 
This new approach is also motivated by its ability to deal with random irregular holes in place of the traditional rectangular masks which may lead to over-fitting and limit the usefulness of the algorithm. 
The main extension introduced by PCNNs is that the convolution is masked and re-normalized in order to be conditioned only on valid pixels. 
This is accomplished with the introduction of the automatic mask update step, which removes any masking where the partial convolution was able to operate on an unmasked value. 
Thus, with this method, the convolutional results depend only on the non-hole regions at every layer such that, given sufficient successive layers, even the largest masks eventually shrink away, leaving only valid responses in the feature map.  

Typically, inpainting in relation to masked CMB maps consists of generating constrained Gaussian realizations to fill in the masked regions \cite{Hoffman1991,Bucher2012}. 
However this approach fails to deal with non-Gaussian components in the map and can be computationally expensive. 
Recently, machine learning algorithms, specifically deep learning using CNNs and GANs, have been increasingly applied to this problem with promising results \cite{Sadr:2020rje,Puglisi:2020deh,Petroff:2020fbf}. 
Nevertheless, all of these approaches have so far only focused on neural networks with pure convolutional layers that fill in square or circular masks. 
In this paper, we decide to adapt the PCNN algorithm developed in \cite{PCNN} to learn the statistical properties of CMB maps and efficiently fill in irregularly masked areas, avoiding biases caused by the initial hole values and the shape of the masks.

%\begin{figure}[ht!]
%    \centering
%    \includegraphics[width=0.7\linewidth, keepaspectratio]{figures/cropped_maps_visual.pdf}
%    \caption[]{In the image it is shown the process to obtain $128\times128$ pixel square CMB chunks used to compose our dataset. The central area of each CMB map, defined as the largest rectangle enclosed in the map, is divided in 10 squares, specifically 5 per row. To conclude, this process is then iterated for all the simulated CMB thermal maps created with \textit{CAMB}.}
%    \label{fig:1}
%\end{figure}

\subsection{Network Architecture}
The proposed model is U-Net \cite{ronneberger2015unet} like and characterized by two main elements: the partial convolutional layer and the loss function. The details are given below and we follow the notation in \cite{PCNN}. 
\subsubsection{Partial Convolutional Layer}
The key element in this network is the partial convolutional layer. Given the convolutional filter weights $\mathbf{W}$, its corresponding bias $b$ and a binary mask of 0s and 1s $\mathbf{M}$; the partial convolution on the current pixel values $\mathbf{X}$ at each layer, is defined as:
        \begin{equation}
             {\scriptstyle x^{\prime}=\left\{\begin{array}{ll}{\mathbf{W}^{T}(\mathbf{X} \odot \mathbf{M}) \frac{\operatorname{sum}(\mathbf{1})}{\operatorname{sum}(\mathbf{M})}+b,} & {\text { if } \operatorname{sum}(\mathbf{M})>0} \\ {0,} & {\text { otherwise }}\end{array}\right.}
        \end{equation}
where $x^\prime$ corresponds to the updated pixel values and $\odot$ is the element-wise multiplication. The scaling factor $\frac{\textbf{sum(1)}}{\textbf{sum(M)}}$ applies some appropriate scaling to adjust for the varying amount of valid (unmasked) inputs.  After each convolution, the mask is removed at the location where convolution was able to condition its output on the valid input. The updated mask pixel values $m^\prime$ are expressed as:
 \begin{equation}
      {\scriptstyle m^{\prime}=\left\{\begin{array}{ll}{1,} & {\text { if } \operatorname{sum}(\mathbf{M})>0} \\ {0,} & {\text { otherwise }}\end{array}\right.}
 \end{equation}
 
\subsubsection{U-Net structure}
The model is characterized by a U-Net architecture where all the convolutional layers are replaced with partial convolutional layers using nearest neighbor up-sampling in the decoding stage. The input images taken in are $128 \times 128$ pixels. 
The coding and decoding stages are made of 7 layers where in the coding phase, the number of filters in each convolution steadily increases from 32 to 128 meanwhile the kernel size decreases from 7 to 3. 
In the decoding stage, the kernel size is kept constant at 3 meanwhile the number of filters in each convolution decreases steadily until the final concatenation with the input image, where it is equal to 3.
The last layer’s concatenation with the original input image and mask serves to allow the model to copy non-hole pixels. Details of the PCNN architecture can be found in the Appendix Tab.\ref{t:1}. 

\subsubsection{Loss Function}
The loss function aims to evaluate both per-pixel reconstruction accuracy and composition. The full form of the loss is taken from \cite{PCNN} and is presented in detail in appendix \ref{sec:loss_terms}.
  \begin{eqnarray}
        \mathcal{L}_{\text{total}} & = & \mathcal{L}_{\text{valid}}+6 \mathcal{L}_{\text{hole}}+0.05 \mathcal{L}_{\text{perceptual}} \nonumber \\
        & & + 120\mathcal{L}_{\text{style}}+0.1 \mathcal{L}_{\text{tv}},
        \label{eq:loss}
    \end{eqnarray}
 where $\mathcal{L}_{\text{valid}}$, $\mathcal{L}_{\text {hole}}$ are respectively the losses on the output for the non-hole and the hole pixels; $\mathcal{L}_{\text {perceptual}}$ is based on ImageNet pre-trained VGG-16 (\textit{pool}1, \textit{pool}2 and \textit{pool}3 layers); $\mathcal{L}_{\text {style}}$ is similar to the perceptual loss where it's first performed an auto-correlation on each feature; and finally $\mathcal{L}_{\text{tv}}$ refers to the smoothing penalty on $R$, where $R$ is the region of 1-pixel dilation of the hole region.

\subsection{Training}

\subsubsection{Dataset}
We use CAMB~\cite{camb2000} to generate 50,000 Gaussian, temperature-only, random CMB realizations for training, validation, and testing. Maps are generated with  Healpix~\cite{healpix2005} $N_{side}=1024$ with the Planck 2018 $\Lambda$CDM parameters~\cite{Aghanim:2018eyx}. 
Images are cut out and projected onto a flat sky, with 128x128, 6.87 arcminute square pixels.
Each image is therefore approximately 14.65 degrees on a side. 
We generate 10,000 images for each of 5 different values of the scalar index $n_s$, between 0.8 and 0.96. We chose the spectral index $n_s$ to be a free parameter because it's the one that, when changed , shows the most significant differences in the CMB patterns. This makes it easier for the program to classify the CMB patches, and for the reader to distinguish visually between the different classes\footnote{$A_s$ also has a similar affect however, as it is a multiplicative scaling parameter, it is unlikely to make the PCNN more generalisable.}.  Each set of 10,000 images is divided into training, validation and test sets, with 7,000, 1,500, and 1,500 images respectively. 
The PCNN architecture takes 3 RGB 8-bit inputs. 
We convert the pixel temperature values to 8-bit RGB values using a linear mapping (e.g. the images are represented as 128x128x3 8-bit integers for a total of 24 bits of effective precision). 
This compression does not impact the results as the true and reconstructed maps are compared post-compression.
The mapping parameters are saved in order to reconstruct the original temperature scale after inpainting.

\subsubsection{Masking
\label{sec:masking}}
The masking is performed through two independent methods.
\begin{itemize}
\item \textbf{Irregular masks}: The first generates masks composed of ellipses, circles and lines in random order, amount and size to cover about 10\% and 25\% of the input $128\times128$ pixel image. We will refer to them as $\mathbf{M_{r}}$ and $\mathbf{M_{R}}$ and they are used, respectively, for the fidelity evaluation and in the training process.  
\item \textbf{Regular masks}
The second method generates centered circular masks of various radius that can cover from 0 to roughly 100\% of the input 128x128 pixel image. This last masking is instead used in the fidelity evaluation both to facilitate the comparison with previous results and to measure the ability of the algorithm to reconstruct the image in terms of the percentage of area covered. We will refer to these masks as $\mathbf{M_{cR}}$ where $R$ is equal to the radius of the circle in pixel units.
\end{itemize}

\subsubsection{Training Procedure}

At first we initialize the weights using the pre-trained VGG 16 weights from ImageNet \cite{VGG-16}. This is done for image classification purposes and recognize the different classes of input CMB patches based on the value of the spectral index, $n_s$. The training is then performed using the Adam optimizer \cite{kingma2017adam}  in two stages. The first stage has a learning rate of $0.0001$ and enables batch normalization in all layers. The second stage has a learning rate of $0.00005$ and only enables the batch normalization in the decoding layers. This is done to avoid the incorrect mean and variance
issues that occur with batch normalization in the masked convolutions (as the mean and variance will be computed for the hole pixels as well) and also helps us to achieve faster convergence.

Overall, the program trained for 29 epochs, with batch size of 5 and respectively 10,000 training steps and 1,500 validation steps. The total computational cost was roughly 100 GPU hours.

\begin{figure*}[hb!]
    \centering
    \includegraphics[width=\textwidth]{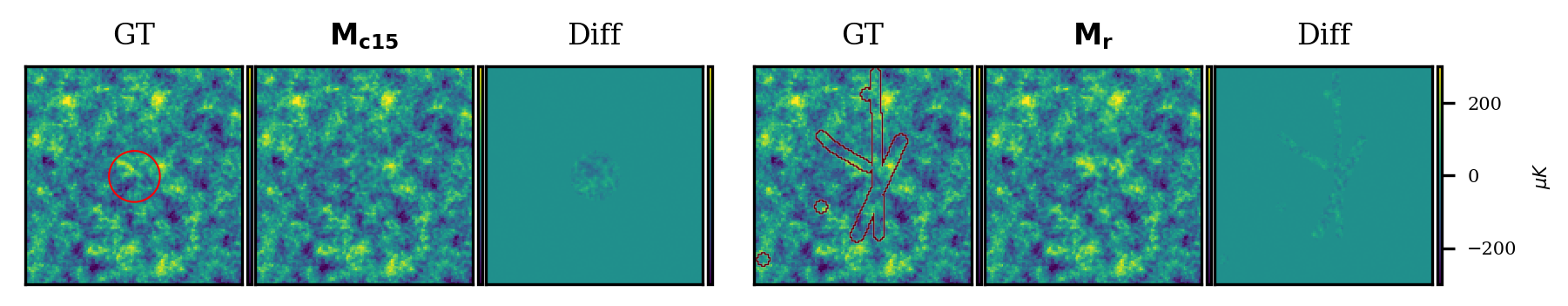}
    \caption{\raggedright The figure shows a ground truth (GT) thumbnail image from the $n_s=0.96$ test set and its corresponding predictions given a circular mask of radius 15 pixels ($\mathbf{M_{c15}}$) and an irregular mask of approximately equal size ($\mathbf{M_{r}}$). For each masking, the difference (Diff) between the inpainted and GT image is also shown. For examples of reconstructed thumbnails from different $n_s$ values and mask size refer to figure \ref{fig:pred_circular_mask} and \ref{fig:pred_random_mask} in Appendix \ref{sup_plots}. }
    \label{fig:PCNN_example}
\end{figure*}

\section{Results}\label{results}

Figures \ref{fig:PCNN_example}, \ref{fig:pred_circular_mask} and \ref{fig:pred_random_mask} show examples of maps extracted
from the test set and reconstructed via the PCNN algorithm for the different masking methods outlined in Section \ref{sec:masking}.

To determine the efficiency of the algorithm in reconstructing the CMB patches, we follow the approach from \cite{Mustafa_2019,aylor2019cleaning,Puglisi:2020deh} and focus on evaluating the ability to replicate the summary statistics of the underlying signal that needs to be reconstructed \cite{Puglisi:2020deh}. 
Specifically, for each $n_s$ value, we compare 500 ground-truth CMB patches from the test set with their corresponding inpainted maps, through two metrics: (1) the pixel intensity distribution; (2) the angular power spectra. Numerically, we perform this comparison by running the Kolmogorov-Smirnov (KS) two-sample test which assesses the likelihood that the distribution of inpainted CMB patches is drawn from the ground-truth distribution.

\subsection{Pixel intensity distribution
\label{results-pd}}

To compute the pixel intensity distribution, all the sample maps are scaled with \texttt{MinMax} rescaling to [-1, 1] as in \cite{Puglisi:2020deh}. 
This is done in order to more easily compare the difference between the ground-truth and
reconstructed maps and Gaussian distributions. 
The left plots in Fig.\ref{fig:evaluation} show the resulting histograms of the pixel intensities for each $n_s$ value, when masked respectively by $\mathbf{M_{c15}}$ (\ref{fig:M_c15}) and $\mathbf{M_{r}}$ (\ref{fig:M_r}). 
The differences between the pixel distributions are negligible and $\lesssim 1 \%$ for both $\mathbf{M_{c15}}$ and $\mathbf{M_{r}}$ and all the $n_s$ values. The resulting $\mathrm{KS}$ test $p$-values are $p>0.999$ for both $\mathbf{M_{c15}}$ and $\mathbf{M_{r}}$, which shows that the PCNN algorithm is able to accurately reproduce the ground-truth CMB patterns.

\subsection{Angular power spectrum
\label{results-ps}}

\begin{table}[ht!]
\centering
    \begin{tabular}{l>{\raggedleft}p{0.14\textwidth}>{\raggedleft\arraybackslash}p{0.14\textwidth}>{\raggedleft\arraybackslash}p{0.14\textwidth}}
    \toprule
    \hline
    $n_s$ &  $\mathbf{M_{c15}}$ & $\mathbf{M_{r}}$ \\
    \hline
    0.80 & $>0.921$ (15) &  $>0.921$ (13) \\
    0.84 & $>0.902$ (14) & $>0.902$ (15) \\
    0.88 & $>0.902$ (15) & $>0.902$ (15) \\
    0.92 & $>0.952$ (09) & $>0.921$ (13) \\
    0.96 & $>0.921$ (11) & $>0.952$ (13) \\
    \hline
    \bottomrule
    \end{tabular}
    \caption[]{\raggedright p-values of the $\mathrm{KS}$ test performed on each multipole bin from the 500 spectra of the ground-truth and inpainted thermal CMB patches. In parenthesis, we report the number of bins with p-value $<0.995$.}
    \label{tab:1}
\end{table}

\begin{figure*}[ht!]
    \centering
    \vspace{-0.35 cm}
    \begin{subfigure}[b]{\textwidth}
    \centering
  \includegraphics[width=\textwidth]{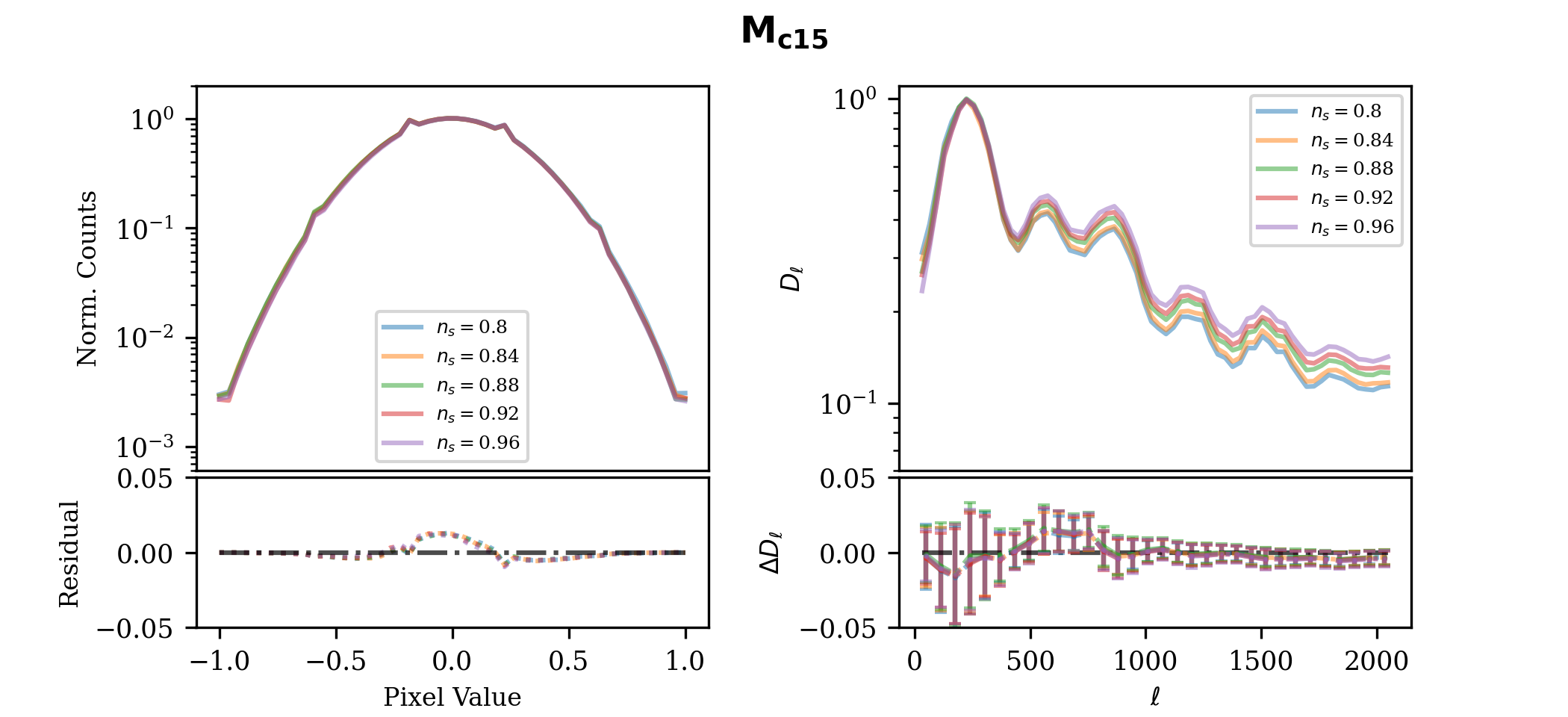}
  \caption{Predictions given a circular mask of radius 15 pixels ($\mathbf{M_{c15}}$).}
  \label{fig:M_c15}
  \end{subfigure}
\hfill
    \begin{subfigure}[b]{\textwidth}
    \centering
  \includegraphics[width=\textwidth]{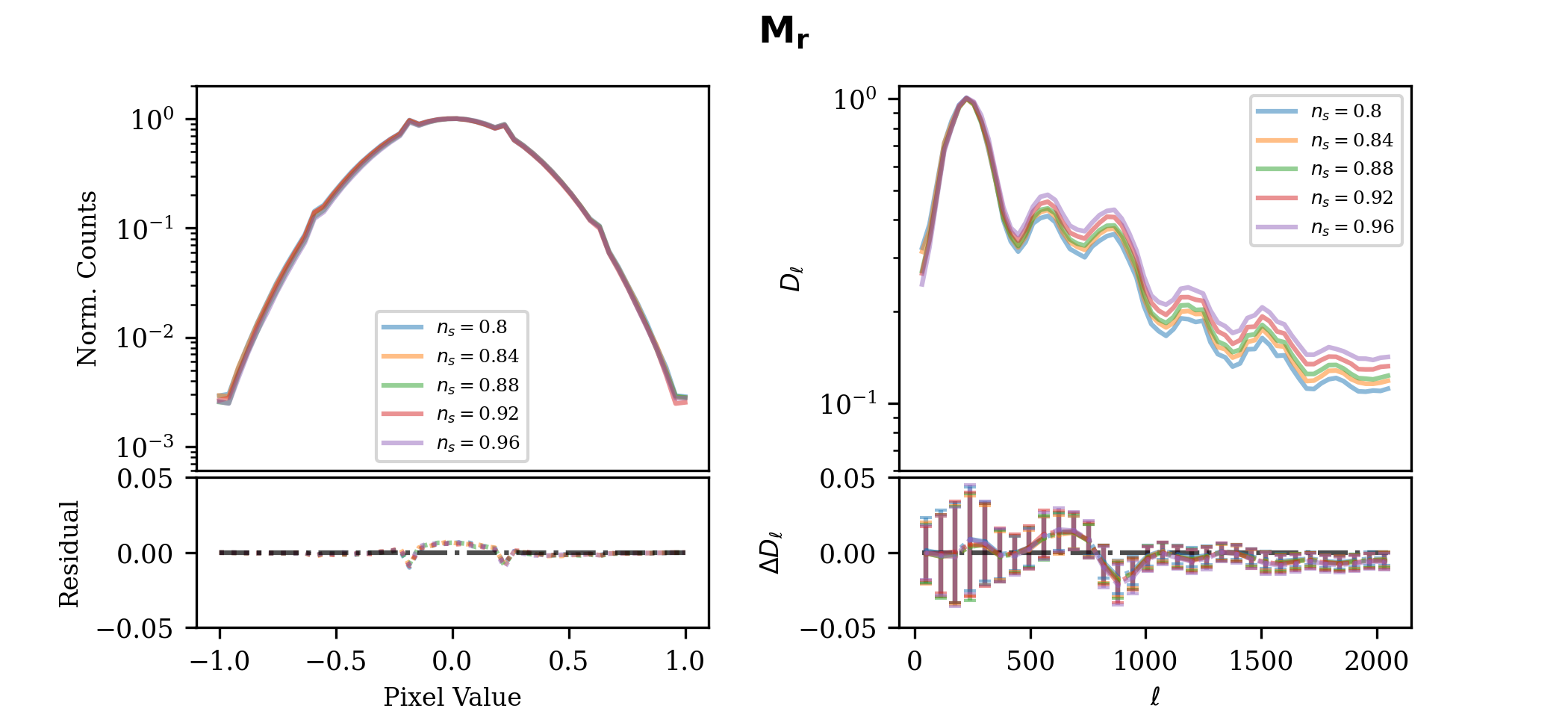}
  \caption{Predictions given an irregular mask of approximately the same size as $\mathbf{M_{c15}}$  ($\mathbf{M_{r}}$).}
  \label{fig:M_r}
  \end{subfigure}
  \caption{\raggedright  The left and right plot of Fig.  \ref{fig:M_c15} and \ref{fig:M_r}  contain respectively the pixel intensity distribution and the average angular power spectrum of 500 inpainted maps with PCNN from the test set for each $n_s$ values considered. The distributions are normalized to 1 and set in log scale. At the bottom of each plot,  the distributions of the residuals between the ground truth and inpainted maps are shown. As expected, for the pixel intensity distributions they are larger in the center rather than on their tails and overall they are always $\lesssim 1 \%$. With regard to the angular power spectra, at high $\ell$ the residuals, though very small, tend to always underestimate the original value. However, they are still consistent with zero as emphasized by the associated error bars (for clarity, the  $\Delta D_{\ell}$ plot was rebinned with $\Delta \ell=64$).}
\label{fig:evaluation}
\end{figure*}

The angular power spectrum in each flat square CMB patch is estimated with \texttt{Pixell}\footnote{\href{https://github.com/simonsobs/pixell}{https://github.com/simonsobs/pixell}}. 
It is binned into equally spaced multipoles with $\Delta \ell= 32$ and maximum multipole $\ell_{\mathrm{max}}=2048$, chosen to correspond roughly to the available sky fraction and \texttt{Pixell} power spectrum accuracy. 
The right plots in Fig.\ref{fig:evaluation} show the resulting average power spectra for each $n_s$ value, when masked respectively by $\mathbf{M_{c15}}$ (\ref{fig:M_c15}) and  $\mathbf{M_{r}}$ (\ref{fig:M_r}).
The power spectra residuals for both $\mathbf{M_{c15}}$ and  $\mathbf{M_{r}}$ are in the same order as the corresponding pixel distribution residuals. At high $\ell$, we note that the residuals are always negative, which implies that PCNN underestimates the original power spectrum at this multipole range. This could be caused by additional smoothing in the inpainted region, leading to reduced small-scale power. That said, it is important to emphasize that the obtained residuals are fully consistent with zero, as shown by the associated error bars, defined as the standard deviation from the mean value of the power spectrum at each $\Delta \ell$ bin. 

To assess more quantitatively the ability to reproduce 
the power spectrum
at a given $\Delta \ell$ bin, we perform the $\mathrm{KS}$ test on the distribution in each bin of the 500 spectra, after having bootstrap resampled it 5000
times.  The obtained $\mathrm{KS}$ test p-values are summarized in Tab.\ref{tab:1}, where we also report in parenthesis the number of bins with p-value $<0.995$. We find that the lowest p-value $>0.902$ with negligible differences between $\mathbf{M_{c15}}$ and $\mathbf{M_{r}}$ and all the $n_s$ values. Hence, we don't find any deviation from the ground truth and we can conclude that the PCNN algorithm is able to coherently reproduce the angular correlations of the underlying CMB signal.
\clearpage

\subsection{Performance in terms of mask size}

\begin{figure}
    \centering
    \includegraphics[width= 0.7\linewidth, keepaspectratio]{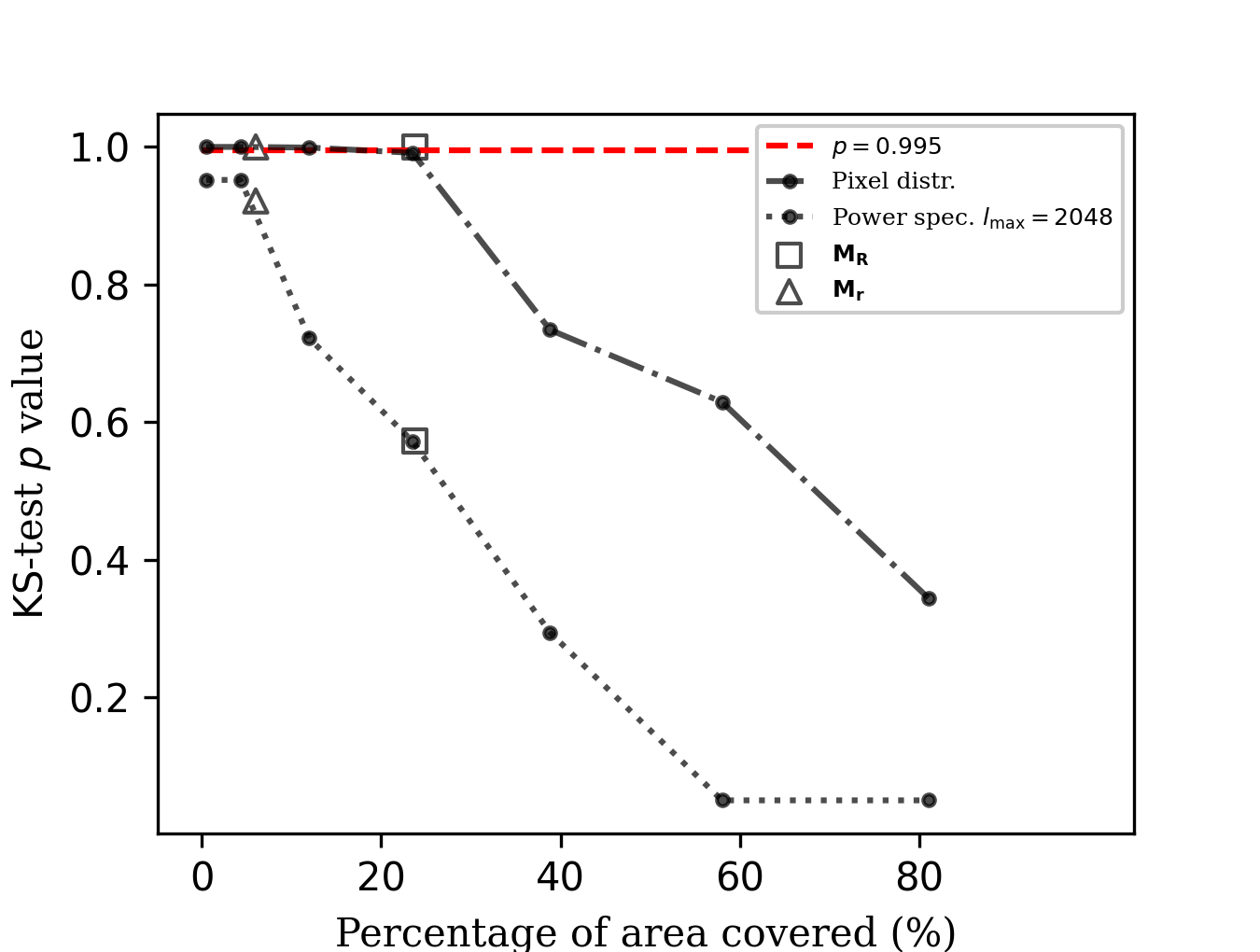}
    \caption{\raggedright The figure presents the worst $\mathrm{KS}$ test $p$-values as a function of the percentage of area covered by circular masks $\mathbf{M_{cR}}$ of radius $R_k=10\cdot k +5$ pixels and $k\in[0,6]$. Respectively, the pixel intensity distribution and the power spectra given $\ell_{\mathrm{max}}=2048$ are the dot-dashed and dotted line. The plot also contains the $\mathrm{KS}$ test $p$-values given irregular masks $\mathbf{M_r}$ (triangle)  and $\mathbf{M_R}$ (square) which present the same values as the corresponding circular masks of equal size, emphasizing the ability of the PCNN algorithm to reproduce ground truth thumbnail images, independently of the mask shape.
    }
    \label{fig:mask_size}
\end{figure}

To assess the performance of the PCNN algorithm in terms of the percentage of CMB patch area covered, we use the circular masks $\mathbf{M_{cR}}$ with increasing radius $R$ from $5$ to $65$ pixels, in steps of $10$ pixels. 
We use the same two metrics as for the fidelity evaluation (pixel intensity distribution and angular power spectrum) and compare 500 ground-truth CMB patches from the $n_s=0.96$ test set  with  their  corresponding  inpainted  maps, for each $\mathbf{M_{cR}}$ masking. 
The choice of the $n_s=0.96$ test over the other $n_s$ values is arbitrary. 
In fact, as presented in subsections \ref{results-pd} and \ref{results-ps}, the obtained $\mathrm{KS}$ test $p$-values are independent of the $n_s$ value under consideration.

The results of this analysis are plotted in Fig. \ref{fig:mask_size} and show  that the worst $\mathrm{KS}$ test $p$-values for the pixel intensity distribution and the angular power spectrum are similar for $\lesssim$ 15\% of masked area and diverge significantly for $\geq$ 20\% of masked area. 
This discrepancy between the obtained $\mathrm{KS}$ test $p$-values for the pixel intensity distribution and the angular power spectrum is most likely caused by the fact that the loss function we use aims to optimize the pixel values of the images. 

Finally, we repeat the same analysis for the irregular masks $\mathbf{M_r}$ and $\mathbf{M_R}$ and note that the shape of the masking has no effect on the algorithm performance. 
In fact, as emphasized in Fig.\ref{fig:mask_size}, the $\mathrm{KS}$ test $p$-values obtained for $\mathbf{M_r}$ and  $\mathbf{M_R}$ are the same as those for the circular masks of equivalent size for all the metrics under consideration.

\section{Summary \& Conclusion}\label{conclusion}

In this paper we have found a novel application for the deep learning algorithm, PCNN: to inpaint masked CMB maps. 
Inpainting CMB maps is normally done on regularly shaped masks. 
There have been a variety of machine learning and other statistical methods to inpaint these types of shapes (circles, squares etc.).
By using PCNNs we can inpaint the CMB maps with regular and irregularly shaped masks.

We train the PCNN by simulating maps of the CMB using known cosmological parameters.
Our method can reconstruct the pixel distribution and the power spectrum of a CMB patch to an accuracy respectively of $<1\%$ and $<5\%$, for masks covering up to $\sim 15\%$ of the image.
Furthermore, for the power spectra we see that the reconstruction error is consistent with zero when the corresponding power spectrum uncertainty is taken into account.
PCNNs work \emph{equally well on regularly and irregularly shaped masks}. 
To quantify how well the PCNN can reconstruct the original images, we perform a KS test and find that the p-value's achieved by the PCNN based on the pixel values of the original and reconstructed images are $\sim 0.999$ for different mask sizes. 
Since in cosmology we are typically interested in the correlation function, as opposed to the amplitude of the pixel values, we also perform a KS test on the true and reconstructed power spectrum, obtaining comparable results. 
This shows that inpainting CMB maps with PCNNs, and computing the power spectra is a viable alternative to standard cut-sky power spectrum estimation using e.g., pseudo-$C_l$ estimators.

The results we have obtained so far have been on a pure CMB map without any noise. 
In reality, any additional sources of noise and other contaminants would have to be included in the forward model used to generate the training dataset. We expect this to have no significant impact on the performance of the proposed algorithm, which is particularly suited to inpaint highly non-Gaussian patterns, as shown already in \cite{PCNN} on images from the ImageNet \cite{VGG-16}, Places2 \cite{places2} and CelebA-HQ dataset \cite{celeb}. 

Overall, our results show that PCNNs can be a powerful method in reconstructing CMB maps to percent level accuracy with irregular masks, in addition to the more common circular and other regular shaped masks. It would be prudent to extend the application of PCNN architectures to inpainting irregular masks on the sphere, as we encounter when performing large sky area CMB analyses. Alongside the map-level reconstruction, we successfully recover the input power spectra from the inpainted maps. It will be interesting to explore further potential biases induced in the power spectrum estimation by the PCNN inpainting process, and studying the use of inpainted power spectrum estimates in cosmological parameter inference. We leave this and attempts to incorporate non-Gaussian foregrounds into our model for future projects.

\section*{Acknowledgments}
MHA acknowledges support from the Beecroft Trust and Dennis Sciama Junior Research Fellowship at Wolfson College.
This project has received funding from the European Research Council (ERC) under the European Union’s Horizon 2020 research and innovation programme (grant agreement No 693024). 
We thank David Alonso, Simone Aiola, Harry Desmond and Mathias Gruber for valuable discussions.

\bibliographystyle{JHEP}
\bibliography{all_active.bib}
\newpage
\appendix
\section{Details of PCNN Algorithm}
\subsection{Network architecture}
\begin{table}[ht!]
\centering
     \begin{tabular}{ m{3.7cm} | m{1.7cm}| m{2cm} | m{1.8cm}| m{1.8cm} | m{2.6cm} }
     Module Name  & Filter Size & \# Channels & Stride Size &
    
    BatchNorm & Nonlinearity \\
     \hline
     PConv1 & $7\times 7$ & 32 & 2 & - & ReLu \\
      PConv2 & $5\times 5$ & 64 & 2 &  - & ReLu \\
       PConv3 & $5\times 5$ & 128 & 2 &  Y & ReLu \\
        PConv4 & $3\times 3$ & 128 & 2 &  Y & ReLu \\
        PConv5 & $3\times 3$ & 128 & 2 &  Y & ReLu \\
        PConv6 & $3\times 3$ & 128 & 2 &  Y & ReLu \\
        PConv7 & $3\times 3$ & 128 & 2 &  Y & ReLu \\
        PConv8 & $3\times 3$ & 128 & 2 &  Y & ReLu \\
     \hline
     NearestUpSample1 & & 128 & 2 & - & - \\
     Concat1 (w/ PConv7) & & 128+128 & & - & - \\
     PConv9 & $3\times 3$ & 128 & 1 & Y & LeakyReLu (0.2) \\
     \hline
     NearestUpSample2 & & 128 & 2 & - & - \\
     Concat2 (w/ PConv6) & & 128+128 & & - & - \\
     PConv10 & $3\times 3$ & 128 & 1 & Y & LeakyReLu(0.2) \\
     \hline
     NearestUpSample3 & & 128 & 2 & - & - \\
     Concat3 (w/ PConv5) & & 128+128 & & - & - \\
     PConv11 & $3\times 3$ & 128 & 1 & Y & LeakyReLu(0.2) \\
     \hline
     NearestUpSample4 & & 128 & 2 & - & - \\
     Concat4 (w/ PConv4) & & 128+128 &  & - & - \\
     PConv12 & $3\times 3$ & 128 & 1 & Y & LeakyReLu(0.2) \\
     \hline
     NearestUpSample5 & & 128 & 2 & - & - \\
     Concat5 (w/ PConv3) & & 128+128 & & - & - \\
     PConv13 & $3\times 3$ & 128 & 1 & Y & LeakyReLu(0.2) \\
     \hline
     NearestUpSample6 & & 128 & 2 & - & - \\
     Concat6 (w/ PConv2) & & 128+64 & & - & - \\
     PConv14 & $3\times 3$ & 64 & 1 & Y & LeakyReLu(0.2) \\
     \hline
     NearestUpSample7 & & 64 & 2 & - & - \\
     Concat7 (w/ PConv1) & & 64+32 & & - & - \\
     PConv15 & $3\times 3$ & 32 & 1 & Y & LeakyReLu(0.2) \\
     \hline
     NearestUpSample8 & & 32 & 2 & - & - \\
     Concat8 (w/ Input) & & 32+3 & & - & - \\
     PConv16 & $3\times 3$ & 3 & 1 & Y & - \\
    \end{tabular}
    \caption{ \raggedright The architecture of the PConv neural network: PConv1-8 are in encoder stage, whereas PConv9-16 are in decoder stage. The BatchNorm column indicates whether PConv is followed
by a Batch Normalization layer. The Nonlinearity column shows whether and what
nonlinearity layer is used. For further details see \cite{PCNN}.}
\label{t:1}
\end{table}
\subsection{Loss Function Terms
\label{sec:loss_terms}}
Given ground truth image $\mathbf{I}_{\mathrm{gt}}$, initial binary mask $\mathbf{M}$, PCNN prediction $\mathbf{I}_{\mathrm{out}}$, we can express the $L^{1}$ losses on the output for the non-hole ($\mathcal{L}_{\text {valid}}$) and hole pixels ($\mathcal{L}_{\text {hole}}$) as:
\begin{equation*}
    \mathcal{L}_{\text {valid }}=\frac{1}{N_{\mathbf{I}_{\mathrm{gt}}}}\left\|M \odot\left(\mathbf{I}_{\text {out }}-\mathbf{I}_{g t}\right)\right\|_{1}; \qquad \mathcal{L}_{\text {hole}}=\frac{1}{N_{\mathbf{I}_{\mathrm{gt}}}}\left\|(1-M) \odot\left(\mathbf{I}_{\text {out }}-\mathbf{I}_{g t}\right)\right\|_{1},
\end{equation*}
where $N_{\mathbf{I}_{\mathrm{gt}}}=C\cdot H\cdot W$ and $C$, $H$
and $W$ are the channel size, height and width of $\mathbf{I}_{\mathrm{gt}}$ \cite{PCNN}. We then define $\mathbf{I}_{comp}$ equal to the raw output image $\mathbf{I}_{out}$ with the non-hole pixels
set to ground truth. The perceptual loss ($\mathcal{L}_{\text {perceptual}}$) computes the $L^{1}$ distances between both $\mathbf{I}_{out}$, $\mathbf{I}_{gt}$ and $\mathbf{I}_{comp}$, after projecting these images into higher level feature spaces using an ImageNet-pretrained VGG-16 \cite{VGG-16}:
\begin{equation*}
    \mathcal{L}_{\text {perceptual}}=\sum_{p=0}^{P-1} \frac{\left\|\Psi_{p}^{\mathbf{I}_{\text {out}}}-\Psi_{p}^{\mathbf{I}_{g t}}\right\|_{1}}{N_{\Psi_{p}^{\mathrm{I} g t}}}+\sum_{p=0}^{P-1} \frac{\left\|\Psi_{p}^{\mathbf{I}_{\text {comp}}}-\Psi_{p}^{\mathbf{I}_{g t}}\right\|_{1}}{N_{\Psi_{p}^{\mathrm{I} g t}}}.
\end{equation*}
$\Psi^{\mathbf{I}_{*}}_{p}$ is the activation map of the $p$th selected
layer given input $\mathbf{I}_{*}$, with $N_{\Psi^{\mathbf{I}_{*}}_{p}}$ equal to its total number of elements. The layers used in this loss are $\textit{pool}1$, $\textit{pool}2$ and $\textit{pool}3$ \cite{PCNN}.
The style loss term ($\mathcal{L}_{\text {style}}$) is equivalent to the perceptual loss on which it's first performed an autocorrelation (Gram matrix) on each feature map \cite{PCNN}.
\begin{eqnarray}
    \mathcal{L}_{\text {style}}=\sum_{p=0}^{P-1} \frac{1}{C_{p} C_{p}}\Big\{\left|| K_{p}\left[\left(\Psi_{p}^{\mathbf{I}_{\text {out}}}\right)^{\intercal}\left(\Psi_{p}^{\mathbf{I}_{\text {out}}}\right)-\left(\Psi_{p}^{\mathbf{I}_{g t}}\right)^{\intercal}\left(\Psi_{p}^{\mathbf{I}_{g t}}\right)\right] \right\|_{1} + \nonumber\\
    \left\|K_{p}\left[\left(\Psi_{p}^{\mathbf{I}_{\text {comp}}}\right)^{\intercal}\left(\Psi_{p}^{\mathbf{I}_{\text {comp}}}\right)-\left(\Psi_{p}^{\mathbf{I}_{g t}}\right)^{\intercal}\left(\Psi_{p}^{\mathbf{I}_{g t}}\right)\right]\right\|_{1} \Big\}, \nonumber
\end{eqnarray}
where $K_p$ is the normalization factor $1/C_pH_pW_p$ for the $p$th selected layer. The final loss term is the total variation loss ($\mathcal{L}_{\text {tv}}$) expressed as:
\begin{equation*}
    \mathcal{L}_{t v}=\sum_{(i, j) \in R,(i, j+1) \in R} \frac{\left\|\mathbf{I}_{c o m p}^{i, j+1}-\mathbf{I}_{c o m p}^{i, j}\right\|_{1}}{N_{\mathbf{I}_{c o m p}}}+\sum_{(i, j) \in R,(i+1, j) \in R} \frac{\left\|\mathbf{I}_{c o m p}^{i+1, j}-\mathbf{I}_{c o m p}^{i, j}\right\|_{1}}{N_{\mathbf{I}_{c o m p}}},
\end{equation*}
where $R$ is the region of 1-pixel dilation of the hole region \cite{PCNN}. 

To conclude, as expressed in eq.\ref{eq:loss}, the total loss ($\mathcal{L}_{\text{total}}$) is a combination of all of the above terms, according to:
\begin{equation*}
     \mathcal{L}_{\text{total}} =  \mathcal{L}_{\text{valid}}+6 \mathcal{L}_{\text{hole}}+0.05 \mathcal{L}_{\text{perceptual}} + 120\mathcal{L}_{\text{style}}+0.1 \mathcal{L}_{\text{tv}}. 
\end{equation*}
were the loss term weights were determined by performing a hyperparameter
search on 100 validation images \cite{PCNN}.
\clearpage
\subsection{PSNR and $\mathcal{L}_{total}$ evolution}
\begin{figure}[hb!]
    \centering
    \includegraphics[width=0.7\linewidth, keepaspectratio]{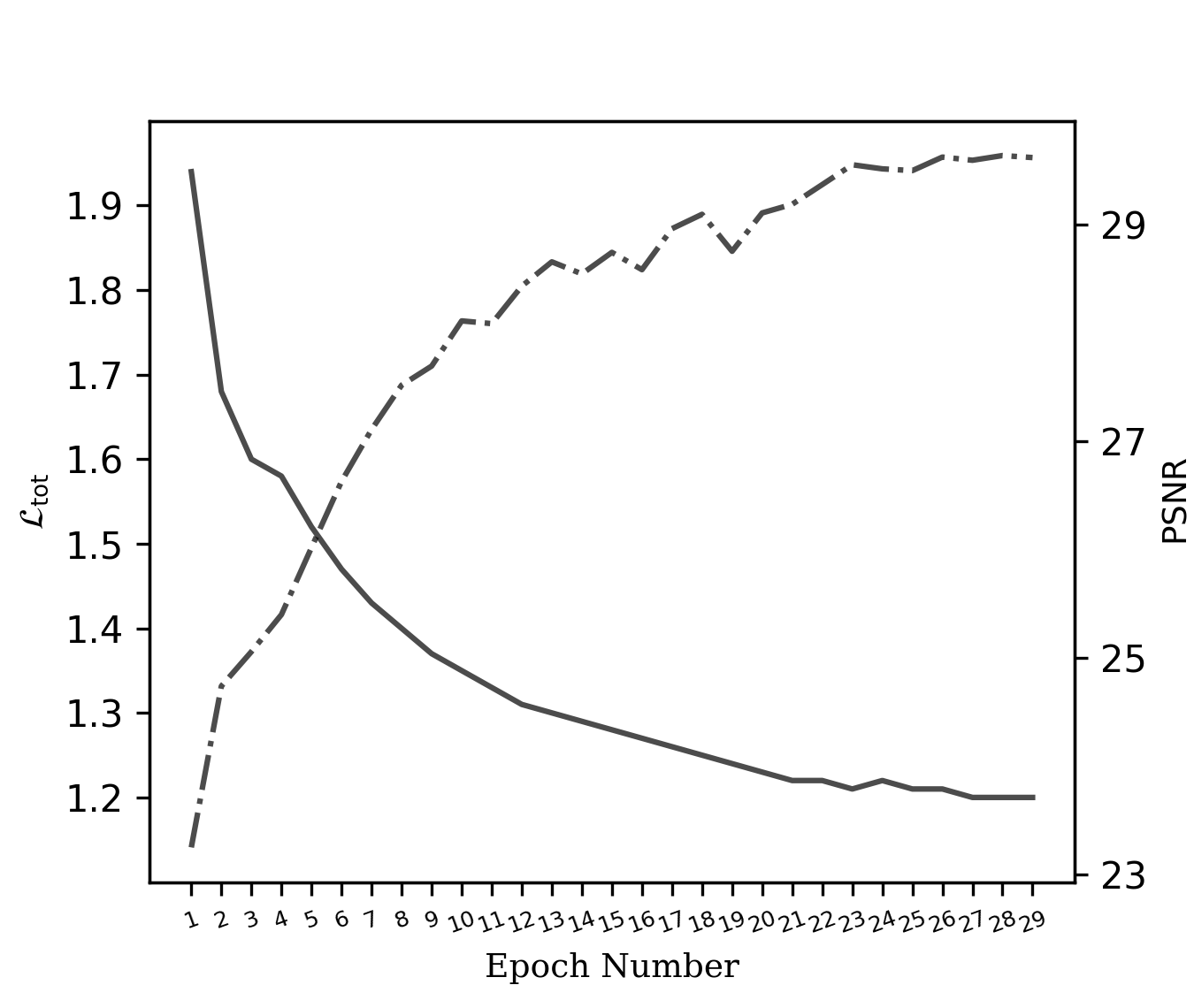}
    \caption{\raggedright The plot shows the evolution of the Peak Signal to Noise Ratio - PSNR (dashed) and the total loss - $\mathcal{L_{\text{total}}}$ (solid) as a function of the epoch number. }
    \label{fig:psnr_ev}
\end{figure}
\clearpage
\section{Supplementary Plots}\label{sup_plots}
\begin{figure}[ht!]
    \centering
    \includegraphics[width=\linewidth, keepaspectratio]{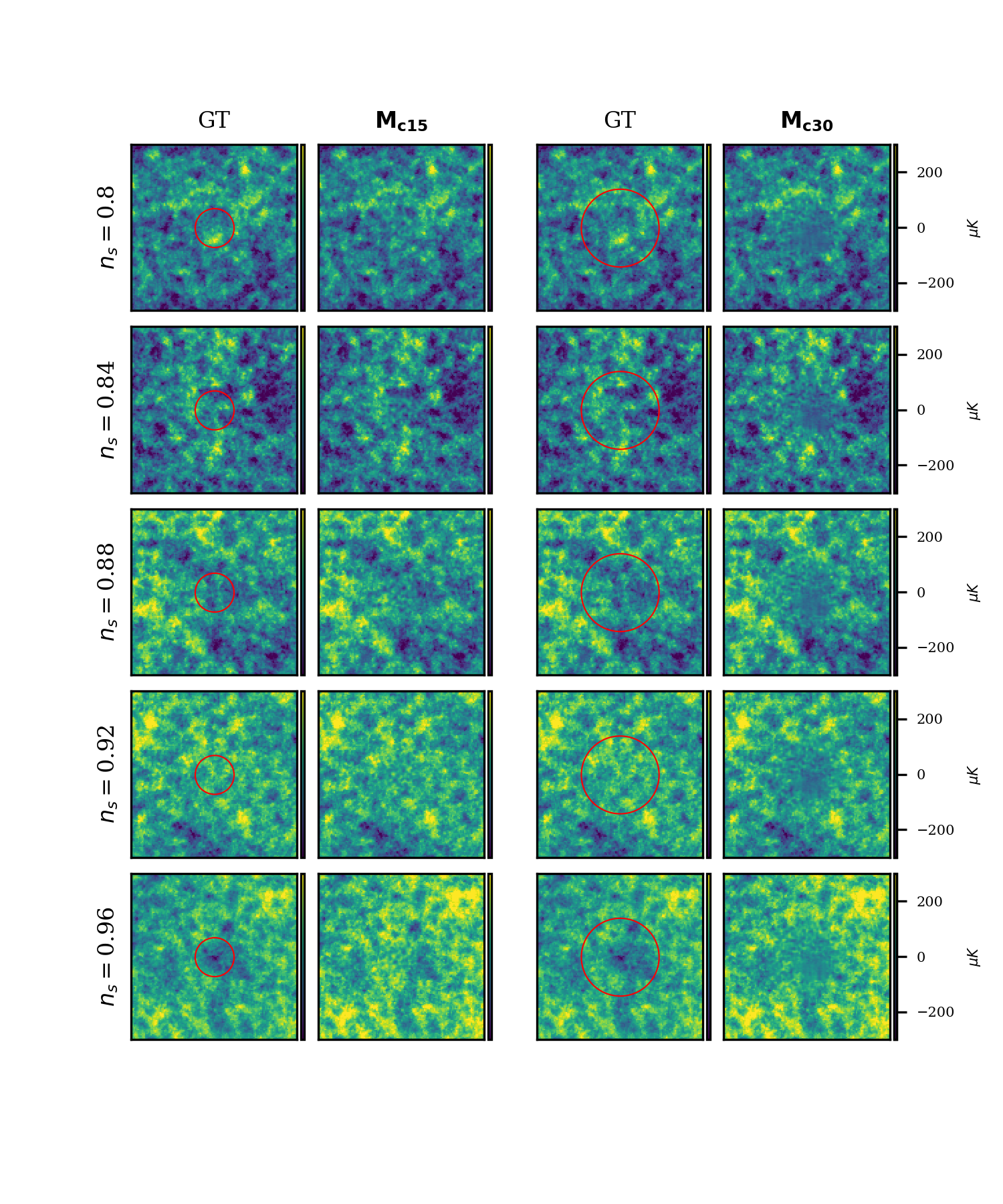}
    \caption[]{\raggedright The Figure shows ground truth (GT) thumbnail images and their corresponding predictions given a circular mask, respectively of radius 15 ($\mathbf{M_{c15}}$) and 30 ($\mathbf{M_{c30}}$) pixels. The masks contours are drawn in red over the GT images. From top to bottom, each row corresponds to a different value of $n_s$. }
    \label{fig:pred_circular_mask}
\end{figure}
\clearpage
\begin{figure}[ht!]
    \centering
    \includegraphics[width=\linewidth, keepaspectratio]{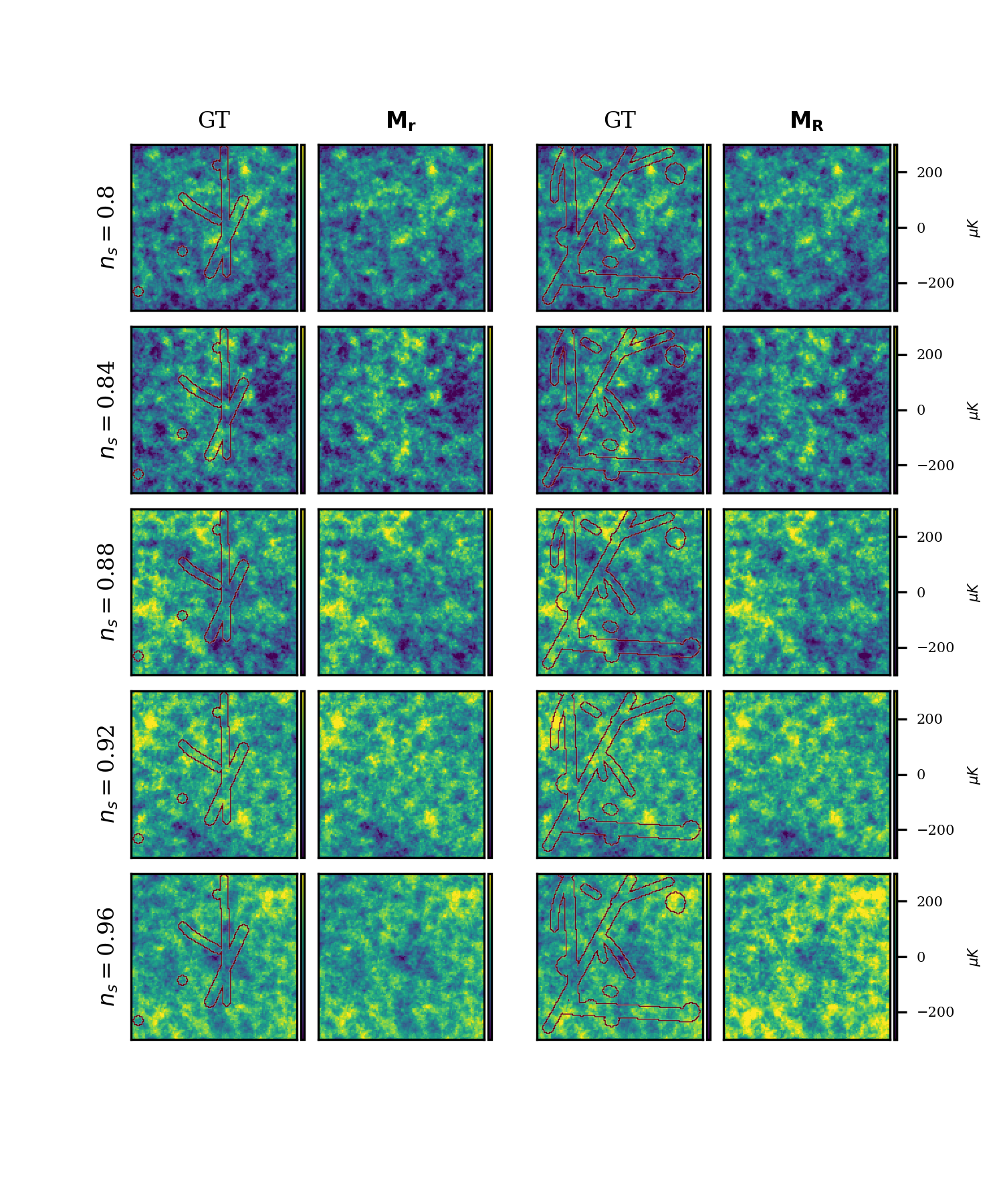}
    \caption[]{\raggedright Similarly to Fig.\ref{fig:pred_circular_mask}, the Figure shows ground truth (GT) thumbnail images and their corresponding predictions. In this case the masks $\mathbf{M_{r}}$ and $\mathbf{M_{R}}$ are irregular and are respectively approximately equal in size to ($\mathbf{M_{c15}}$) and ($\mathbf{M_{c30}}$). As in fig.\ref{fig:pred_circular_mask}, the masks contours are drawn in red over the GT images and from top to bottom, each row corresponds to a different value of $n_s$.
    }
    \label{fig:pred_random_mask}
\end{figure}

\end{document}